\documentclass[prd,nofootinbib,showpacs,preprintnumbers,amssymb]{revtex4} 
\usepackage{graphicx}
\usepackage{epsfig}
\usepackage{dcolumn}
\usepackage{bm}
\usepackage{amsmath}
\def\gsim{\lower0.5ex\hbox{$\:\buildrel >\over\sim\:$}}
\def\lsim{\lower0.5ex\hbox{$\:\buildrel <\over\sim\:$}}

\def \n{\noindent}

\let\d=\delta

\let\m=\mu\let\n=\nu
\let\t=\tau

\newcommand{\be}{\begin{equation}}
\newcommand{\ee}{\end{equation}}
\newcommand{\bea}{\begin{eqnarray}}
\newcommand{\eea}{\end{eqnarray}}

\newcommand{\del}{\partial}

\newcommand{\nbox}{{\,\lower0.9pt\vbox{\hrule \hbox{\vrule height 0.2 cm
\hskip 0.2 cm \vrule height 0.2 cm}\hrule}\,}}

\begin{document}

\preprint{UCI-TR-2010-03, UH-511-1148-10, pi-cosmo-175}

\title{
Constructing Infrared Finite Propagators in Inflating Space-time}

\author{Jason Kumar$^a$}
\email{jkumar@hawaii.edu}
\author{Louis Leblond$^b$}
\email{lleblond@perimeterinstitute.ca}
\author{Arvind Rajaraman$^c$}
\email{arajaram@uci.edu}
\affiliation{
$^a$ Department of Physics and Astronomy, University of Hawai'i, Honolulu, HI
96822, USA\\
$^b$ Perimeter Institute for Theoretical Physics,
  Waterloo, Ontario N2L 2Y5, Canada
  \\
  $^c$Department of Physics and Astronomy, University of California,
Irvine, CA 92697, USA}

\date{\today}

\begin{abstract}
The usual (Bunch-Davies) Feynman propagator
of a massless field is not well defined in an expanding universe due
to the presence of infrared divergences. We propose a new propagator which yields infrared finite answers
to any correlation function. The key point is that in
a de Sitter spacetime there
is an ambiguity in the zero-mode of the propagator.  This
ambiguity can be used to cancel the apparent divergences which
arise in some loop calculations in eternally (or semi-eternally)
inflating spacetime. We refer to this process as zero-mode modification.
The residual ambiguity is fixed by observational
measurement.  We also discuss the application of this method to
calculations involving the graviton propagator.
\end{abstract}


\maketitle
\section{Introduction}
Improvements in cosmological observations
may soon lead to the first detection of non-Gaussianity in the cosmic microwave
background~\cite{Komatsu:2008hk} or in the large scale structure of the
universe \cite{Slosar:2008hx}.
These non-Gaussianities could have a primordial origin in theories where the inflaton
has self interactions or interactions with other light fields (see ~\cite{Bartolo:2004if}
for a review). As such there has been much recent interest in revisiting cosmological
perturbation theory beyond leading order. In addition to non-Gaussianities, the inclusion of
interactions of the inflaton naturally leads one to wonder about loop corrections to other
observables, such as the power spectrum.

From dimensional analysis one might expect these loop corrections to be very small, and many
recent calculations have confirmed this expectation in several different settings \cite{Weinberg:2005vy}.
Nevertheless, there is some expectation that loop corrections may be bigger than naive dimensional analysis would predict.
The reason is that in slow roll inflation, the
inflaton is almost massless (i.e.~its mass is much lower than the Hubble scale during
inflation). If the inflaton is treated as a massless field,
loop corrections to correlation functions
are afflicted with
infrared (IR) divergences due to massless particles
propagating in loops, and these make the
loop diagrams formally infinite as compared to the
tree level contributions. This would lead to the breakdown of perturbation theory,
and the tree level calculation would be rendered untrustworthy.
A resolution
of these IR divergences is essential to extract proper predictions\footnote{Note that in single field inflationary models, the
curvature perturbation
defined on three dimensional hypersurfaces only has derivative interactions and loop
corrections are IR finite. We do not expect this to be generally true in multi-field
models of inflation or when computing quantities other than correlation functions of
the curvature perturbations.}.

These divergences have been investigated by many authors \cite{Allen:1987tz, Tsamis:2005hd, Boubekeur:2005fj, Boyanovsky:2005px, Urakawa:2009my}.
In principle, one may avoid the issue of these divergences by including
deviations from exact masslessness and exact de Sitter geometries.
For example, if the inflaton has a mass, the IR divergences will be cut off
(indeed, since the mass is perturbatively renormalized, it requires
a severe fine tuning to keep the field truly massless). It has been argued
that in $\lambda \phi^4$ theory, a mass is dynamically generated
\cite{Starobinsky:1994bd, Riotto:2008mv, Burgess:2009bs}.
The tilt in the potential in slow-roll models can also lead to a cutoff of the divergences.
Finally, if inflation only occurs over a finite period, the boundary
conditions in the far past also modify the behavior of the low momentum modes
and cut off the divergences \cite{Vilenkin:1982wt}.

For these reasons, it might appear natural to treat the IR divergences of the massless field
as unphysical, and to assume that one or more of these mechanisms
is responsible for obtaining finite results. This is unsatisfactory for
a few reasons. In the first place, it is perfectly possible to consider
a massless scalar field that is not the inflaton, and which can have
an arbitrary potential. There appears to be no obvious reason
that the quantum field theory of such a scalar
should break down in the exact de Sitter geometry.
Secondly, even if the IR divergences
are cut off by these physical corrections, the loop corrections may be large,
and we will need to understand their effects. Finally, gravitons in
de Sitter space also have propagators similar to
those of massless scalar fields. Since their mass is not perturbatively renormalized,
we are forced to deal with the divergences when we calculate corrections
to observables which involve graviton loops.

Here, we shall reexamine these divergences in the context of
a massless scalar field theory in a (flat) Friedman-Robertson-Walker (FRW) space-time with constant Hubble rate
(this scalar is {\it not}  the inflaton). This spacetime is a Poincar\'e patch of exact de Sitter;
it is locally identical to de Sitter but has only  six global Killing vectors.
We will ignore the
backreaction of the scalar on the metric (implicitly assuming
that $M_{Pl}\rightarrow \infty$).
We argue that the resolution to the IR divergences involves a correct understanding of what the
Feynman propagator is in an inflating spacetime.
In fact, just as in \cite{Kirsten:1993ug}, we find that
 a proper treatment of the zero momentum mode removes this
divergence. The main difference between the FRW case and the de Sitter case (see~\cite{Kirsten:1993ug}) is that
the spectrum is continuous, and some care must be taken to correctly
change only the zero mode.  The end result is IR finite but will in general depend on time
(which is fine for a FRW
background).
We then show that this
modification is compatible with other approaches to treating the IR divergences. However,
our method is more direct and is easier for computational purposes.

In section 2, we describe the correct treatment of the zero mode in the propagator, and
in section 3 we illustrate with an example of a loop calculation.
In section 4, we consider the graviton case.
As in the scalar case, the graviton propagator must be
modified at tree level. However,this modification applies
only to the graviton zero mode, which is a pure gauge mode. By
choosing an appropriate gauge, the IR divergences can be gauged away (as argued to
be true in \cite{Higuchi:1985ad, Allen:1986ta} (see also \cite{Higuchi:2001uv, Higuchi:2002sc} for
more recent discussions). This provides an IR finite method for computing graviton
loop corrections.  Our conclusions are presented in section 5.

\section{Defining the Feynman Propagator}

We will consider the action for a scalar in a FRW background
\bea
S=\int dt d^3x \sqrt{g}\ \left(\frac{1}{2}g^{\m\n}\del_\m \phi\del_\n \phi
-V(\phi)\right)
\eea
where
\bea
ds^2=g_{\m\n}dx^{\mu} dx^{\nu} \equiv a^2(\tau)(- d\tau^2 +d\vec{x}^2)
\eea
and we take $a(\tau)=-\frac{1}{H\tau}$.
In the non-interacting limit with a pure cosmological constant $V(\phi)=V_0$,
the classical equation of motion for the mode functions can be solved to yield
\bea\label{BD}
\phi(\vec{k},\t)=-i{H\over \sqrt{2} k^{3/2}}(1-ik\tau) e^{ik\tau}
\eea
where we have chosen the usual Bunch-Davies vacuum and $H = \frac{a'}{a^2} = \frac{1}{M_{Pl}}\sqrt{\frac{V_0}{3}}$.
The Feynman propagator is then
\bea\label{BDprop}
G_F(\vec{x},\t ;\vec{x}',\t')&=&\langle 0|T\phi(\vec{x},\t)\phi(\vec{x}',\t')|0\rangle
\nonumber\\
&=&
\int {d^3k\over (2\pi)^3}{H^2\over  2k^{3}} (1+ k^2 \t \t' +\imath k|\t -\t'|)
e^{-\imath k|\t -\t'|} e^{\imath \vec{k}(\vec{x}-\vec{x}')}\; .
\label{IRdivprop}
\eea
When interactions are turned on, there will be quantum corrections to this correlation function.
Since $\langle\phi(\vec{k},\t)\phi(\vec{k},\t')\rangle \propto k^{-3}$ for small $k$,
one would expect loop corrections to include a contribution from the
low-momentum region of the form $\int {d^3k\over k^3}$, leading to a
logarithmic IR divergence.
In fact the expression (\ref{IRdivprop}) is infinite even for finite scale factor
and points which are at finite separation (due to the divergence at $k=0$).
It has been suggested~\cite{Starobinsky:1994bd, Riotto:2008mv, Burgess:2009bs} that this divergence can be cured by nonperturbative
effects, which can generate a mass for the otherwise massless scalar field.
However, this non-perturbative contribution can always be canceled by
a suitable finite counterterm,  again yielding a massless particle.  We can therefore
assume that our renormalization conditions have been chosen to give us a truly
massless field $\phi$, and in this case we still must ask how this IR divergence is
to be treated.

Intuitively, though, it does not seem that the divergences in
(\ref{IRdivprop})
should be physical. For practical purposes, the largest observable scales today exited the inflationary
Hubble patch  60 e-folds before the end of inflation.  The IR divergence
in (\ref{IRdivprop}) arises from modes which left the horizon long before that.
As these modes themselves are not physically distinguishable from a constant zero-mode, it would
seem possible to treat virtual propagation of these modes in a way which removes the
divergence.

To address this problem, we note that
one can add a homogeneous solution to any inhomogeneous solution
to obtain another solution to the inhomogeneous wave-equation; these correspond
to propagators with different boundary conditions.
The time-ordering of the Feynman propagator defines the boundary conditions:
positive energy modes propagate forward in time, while
negative energy modes propagate backward in
time.
However, in the massless case, there is a zero frequency mode.
The boundary condition on this mode is not
set by the causality condition, and is an ambiguity in the Feynman
propagator.
This is perhaps easier to see in position space where
the propagator
satisfies the inhomogeneous wave equation
\bea\label{eom}
\del^\m  \del_\m G_F(\vec{x},\t;\vec{x}' ,\t')=-i\d^3(\vec{x}-\vec{x}')\d(\t-\t')\label{propeqn}\; .
\eea
This has a constant homogeneous solution, and hence the
propagator can be shifted by a constant.
This constant piece can be thought of as a shift to the 2-point correlation function, and
contributes to the Feynman propagator (we note that since the
retarded propagator is
$G_R(\vec{x},\t;\vec{x}' ,\t') =\theta(\t-\t') \langle [\phi(\vec{x},\t) ,\phi(\vec{x}' ,\t')]\rangle$,
the constant shift cancels in the commutator, as required by causality.)

Note that the propagator for a massless field in Minkowski space suffers from a precisely
analogous ambiguity; there, however, the requirement from cluster decomposition that the propagator
fall off at large distance eliminates the ambiguity. In de Sitter
space (or FRW) such a requirement cannot be imposed; the propagator necessarily grows
at large distances.

We can therefore add a (possibly divergent) constant to the
propagator to make it finite. We modify the propagator
by
putting an IR cutoff at $k=\mu$ (and taking $\mu \rightarrow 0$). The
finite propagator is defined to be
\bea
\label{ZMRF1}
G^{\rm fin}_F(\vec{x},\t,\vec{x}',\t')=\lim_
{\mu\rightarrow 0}
\int_{\mu} {d^3k\over (2\pi)^3}{H^2\over  2k^{3}} (1+ k^2 \t \t' +\imath k|\t -\t'|)
e^{-\imath k|\t -\t'|} e^{\imath \vec{k}(\vec{x}-\vec{x}')}
+f\left({k_{IR}\over \mu}\right)
\eea
where we have added a constant $f\left({k_{IR}\over \mu}\right)$. To cancel the leading divergence as $\mu\rightarrow 0$
we take $f\left({k_{IR}\over \mu}\right)=- \left(\frac{H}{2\pi}\right)^2 \ln{k_{IR}\over \mu}$
(here we have introduced a new comoving scale $k_{IR}$.)
The propagator in momentum space then can be written as
\bea
\label{ZMRmom}
G^{\rm fin}_F(k,\t,\t')=\lim_{\mu\rightarrow 0} \left[{H^2\over 2k^{3}}
(1+ k^2 \t \t' +\imath k|\tau -\tau'|)
e^{ik|\tau-\tau'|}\theta(k-\mu)
+f\left({k_{IR}\over \mu}\right)(2\pi)^3\d^3(\vec{k})\right]\; .
\label{modifiedprop}
\eea
It is easy to see that the Feynman propagator is finite. Indeed, all calculations
in perturbation theory will now yield results free of IR
divergences. We refer to this procedure as {\it zero-mode
modification}. Note that while the divergences have canceled, we have introduced a
scale $k_{IR}$ which is a finite ambiguity in the propagator; this ambiguity will need to
be fixed by an experimental measurement.
A similar analysis applies for the two point correlation function. The usual
formula for this quantity is
\bea
\label{pos1}
G_0(\vec{x},\t,\vec{x}',\t')=\langle\phi(\vec{x},\t)\phi(\vec{x}',\t')\rangle=
\int\frac{d^{3} k}{(2\pi)^3} {H^2\over 2k^{3
}}(1+ik\tau)(1-ik\tau')
e^{-i\vec{k}(\vec{x}-\vec{x}')}e^{ik(\tau-\tau')}\; .
\eea
The zero-mode modified correlation function is
 defined analogously to be
\bea
\label{ZMRpos1}
G^{\rm fin}_0(\vec{x},\t,\vec{x}',\t')=\lim_
{\mu\rightarrow 0}\int_{\mu} \frac{d^{3} k}{(2\pi)^3} {H^2\over 2k^{3
}}(1+ik\tau)(1-ik\tau')
e^{-i\vec{k}(\vec{x}-\vec{x}')}e^{ik(\tau-\tau')}
+f\left({k_{IR}\over \mu}\right)\;
\eea
and is also free of divergences.

It may appear that this method of treating the zero-mode is
discontinuous; one adds a constant to the propagator of the zero-mode, while
leaving undisturbed the propagator of modes with arbitrarily small
wavenumber.  However, this discontinuity cannot be observed, as any
physical experiment
can only probe wavelengths
smaller than the size of the observable universe.  One cannot distinguish
the manner in which small wavenumber modes are treated from the manner of
treatment of the
zero-mode.

The effect of zero-mode modification is to effectively remove the
contributions from modes below $k_{IR}$; it therefore has some similarities to
imposing a hard cutoff at that scale (it is important to note, however, that
our scale $k_{IR}$ does not correspond to an actual cutoff on the
mode integration.  In our formalism, all modes are integrated over,
and the effective scale $k_{IR}$ is independent of this limiting process.)
However, when we consider higher point functions,
 the integrand of the momentum integration will have additional
terms, i.e. it will be of the
form ${1\over k^3}(1+a_1 k+a_2 k^2 +...)$.
The IR singularity from the leading term is
still canceled by
zero-mode modification, but subleading terms will yield a finite contribution
from the modes $k < k_{IR}$.  This contribution would not be present in the
case of
a hard-cutoff at $k_{IR}$, and this feature
could potentially distinguish these two approaches.

Note that in order for eq.~(\ref{ZMRmom}) to be a solution of eq.~(\ref{eom}), $k_{IR}$ and $\mu$ must be comoving scales
and thus independent of time. UV regularization on the other hand needs to be done with a UV cutoff fixed
in physical scale ($\Lambda_{UV} = k_{UV} a(\tau) $) and therefore
the 
coincidence limit of the two-point function  depends on time (as in~\cite{Vilenkin:1982wt})
\be
\label{timegrowth}
G_F^{\rm fin}(\vec{x},\tau, \vec{x},\tau) \sim \ln \frac{a(\tau)}{a_0}
\ee
with an arbitrary constant $a_0$.

This formula also offers an alternative way of understanding our modified propagator.
Equation (\ref{timegrowth}) implies that the variance of a massless field
grows with time.
This fact is most easily understood in  stochastic
inflation.    In this approach, one
treats all modes of  wavelength larger than some cutoff
(typically the scale of the observable universe) as part of a zero
mode.
As each mode at the scale of the horizon freezes out,
it shifts the value the stochastic field by a fixed magnitude,
but with a random phase.  The value of the field thus executes a random walk,
which leads to a variance which grows linearly with time.
If inflation is taken to begin at time
$t=-\infty$, then for any finite time, we get an infinite variance.
Alternatively~\cite{Kirsten:1993ug}, one can quantize the zero mode of the field separately,
such that vacuum state is the product of a Hilbert space (for the zero mode) and a Fock space.
In this formalism, the zero mode is a free particle. The de Sitter invariant
solution is then to put the zero mode into an eigenstate
of the conjugate momentum operator. This again leads to
an infinite variance (since by the uncertainty relation, the value of the field
is arbitrary.) This result is naturally consistent with the stochastic interpretation;
in the stochastic interpretation, the random walk of the scalar field causes the variance
to increase with time, and this can only be de Sitter-invariant if the variance is
always infinite.

In the zero-mode modified propagator, this divergence is not present. Our propagator should be therefore
interpreted as having boundary conditions where the variance is finite at finite times, but is not
finite as $t=-\infty$. Alternatively, we are considering a state where
the wavefunction of the zero mode is not infinitely delocalized.
Our choice breaks de Sitter invariance; by taking a limit
where the constant $k_{IR}$ in (\ref{IRdivprop}) becomes zero,
we can restore de Sitter invariance at the cost of reintroducing the
IR divergences.

It is also possible to think of our zero mode modification as a limiting case of the more standard scheme of
modifying the Bunch-Davies vacua.  Indeed, as considered in~\cite{Ford:1977in}, one way to obtain IR finite
answers in a FRW spacetime is to modify the wave functions to
\be
\phi(\vec{k},\tau) = c_1(k) \tau^{1/2} H_{3/2}^{(1)} (-k\tau) + c_2(k) \tau^{1/2} H_{3/2}^{(2)}(-k\tau)
\ee
where $H_{3/2}^{(1,2)}$ are the Hankel functions. For appropriate choices of $c_1,c_2$ e.g. \bea
c_1 = k^{-p}\qquad
c_2 = \left(k^{-2p} + \frac{3\pi}{4}\right)^{1/2}
\eea
the divergences can be removed. The Bunch-Davies vacuum (\ref{BD}) corresponds
to $c_1 = 0$ and $c_2 = \sqrt{3\pi \over 4}$.  Zero-mode modification
corresponds to the limit $p\rightarrow \infty$, which
 goes over to the Bunch Davies vacuum
for all nonzero $k$. For $k=0$, the limit is singular; this corresponds again
to the ambiguity in the zero mode.

\section{Application to a Loop Calculation}

We now apply this modified propagator to a simple loop calculation.
Consider a function $N(\phi)=N_0+N_1\phi+N_2\phi^2+...$. As our notation suggest, this function could be related
to the number of e-folds which in turn determines the scalar curvature perturbation.
The two-point correlation function is
\bea
\langle N(x,\t)N(y,\t')\rangle & = &N_0^2+N_1^2\langle \phi(x,\t)\phi(y,\t')\rangle
+N_2^2\langle \phi^2(x,\t)\phi^2(y,\t')\rangle+\cdots\nonumber\\
& = & N_0^2+N_1^2G_0(x,t;y,t')
+N_2^2(G_0(x,t;y,t'))^2+\cdots
\eea
where $G_0$ is the 2-point correlator.
When this is evaluated in momentum space, we find
\bea\label{cloopeq}
\langle N(\vec{k},\t)N(\vec{k},\t')\rangle & =&N_0^2+N_1^2\langle \phi(\vec{k},\t)\phi(\vec{k},\t')\rangle
+N_2^2\int \frac{d^3k'}{(2\pi)^3} \langle \phi(\vec{k}',\t)\phi(\vec{k}-\vec{k}',\t)
\phi(\vec{k}',\t')\phi(\vec{k}-\vec{k}',\t')\rangle+\cdots\nonumber\\
&=& N_0^2+N_1^2 G_0(\vec{k}, \t ,\t')
+N_2^2\int \frac{d^3k'}{(2\pi)^3}  G_0(\vec{k}', \t ,\t')   G_0(\vec{k}-\vec{k}', \t ,\t')
+\cdots
\eea
The last term can be represented as a loop diagram.
If one uses the expression (\ref{pos1}) for the correlation function,
 the loop integral will exhibit an IR divergence.  In
particular, if the lower limit of integration is some cutoff scale
$\mu$, then the divergence is manifested through the
appearance of $\ln \mu$, which diverges in the $\mu \rightarrow 0$ limit.

We now show that by using the zero-mode modified correlation function (\ref{ZMRpos1}), there is
no IR divergence.
An infrared divergence in the integral in (\ref{cloopeq}) arises if $k' \sim 0$ or
$k' \sim k$, in which case one factor $G_0$ is potentially
divergent. The integral converges very quickly for
$k' \gg k$ and so the incoming momenta serve as a UV cutoff for this loop diagram.
Without loss of generality,
consider
the integrand in the limit $k' \sim 0$, and integrate over the modes between
zero and $k$.  We then find that the relevant  integral is (after zero-mode modification)
\bea
N_2^2\int_0^{k} \frac{d^3k'}{(2\pi)^3} G^{\rm fin}_0(\vec{k}', \t ,\t')   G^{\rm fin}_0(\vec{k}-\vec{k}', \t ,\t')
&\sim&
N_2^2 G^{\rm fin}_0(\vec{k}, \t ,\t') \int_0^{k} \frac{d^3k'}{(2\pi)^3} G^{\rm fin}_0(\vec{k}', \t ,\t')\; .
\eea
After zero-mode modification, the integral over all modes with
momentum $< k_{IR}$ cancels against the zero mode, and we get the contribution
\bea
N_2^2 G^{\rm fin}_0(\vec{k}, \t ,\t') \int_{k_{IR}}^{k} \frac{d^3k'}{(2\pi)^3} G^{\rm fin}_0(\vec{k}', \t ,\t')
\sim  N_2^2 G^{\rm fin}_0(\vec{k}, \t ,\t') \left(\frac{H}{2\pi}\right)^2 \ln\frac{k}{k_{IR}} \; .
\eea
We see that the contribution to the integrand from the putatively
divergent regions is canceled.  The integral over the
remaining modes yields a dependence on $\ln (k/k_{IR})$, but there
is no divergence.

Note also that a shift in the quantity $k_{IR}$, shifts the correlation function
by a term proportional to $G^{\rm fin}_0$, and can be absorbed by a change in $N_1$. This is
analogous to a renormalization group flow, where the shift in the scale leads to a shift
in the coupling constants. In any experimental measurement, we can choose $k_{IR}$ to be
the scale of observation;
for example, in \cite{Kumar:2009ge}, one was interested in temperature correlation function in the CMB on large scales
as measured by WMAP and Planck and
hence $k_{IR}$ was chosen at a scale of order of  $H_0$, the Hubble radius today.  The ambiguity
in the propagator is thus absorbed into coefficients like $N_1$, which must be measured.
More generally, a smoothing (window) function that depends on some experimental scale must be used when comparing to any experiments;
this smoothing procedure ends up setting the arbitrary scale $k_{IR}$ to be the observational smoothing scale $1/H_0$ in many calculations.
Not every physical quantity needs to be smoothed;
gravitational waves, for example, are not subtracted against any background zero mode and, as we will discuss below,
our modified propagator is quite useful in this case.

Up to now, we have been looking at a free field and modifying the propagators at tree-level. Interactions can introduce
further infrared divergences, which will need to be canceled by modifying
$f({k_{IR}\over \mu})$
at higher orders. For example, the two point function $\langle \phi(\vec{k},\tau_1) \phi(-\vec{k}, \tau_2) \rangle $
receives a series of corrections in an interacting theory
of the form
\bea
\int d\tau_1 d\tau_2G^{\rm fin}_0(\vec{k},\tau_1,\tau') \Sigma(k,\tau_1,\tau_2)G^{\rm fin}_0(\vec{k},\tau',\tau_2)
\eea
where $\Sigma$ is a loop diagram. In general, $\Sigma$ will contain terms of the
form $\ln (k_{IR}\tau_1)$ or $\ln (k_{IR}\tau_2)$, which will lead to apparent divergences when
integrating over $\tau_1,\tau_2$.

However, the integrand is cut off for $| k\tau' | \gg 1 $ by the oscillatory terms
in $G_0(\vec{k},\tau_1,\tau')$. This removes the divergence for small $\tau_1,\tau_2$, but leads to terms
of the form $\ln ({k_{IR}\over k})$.
Such terms are finite if $k> k_{IR}$ but significantly modify the propagator for $k \ll k_{IR}$.
To cancel the contribution from these modes, $f({k_{IR}\over \mu})$ will have to be adjusted
order by order in perturbation theory.

\section{The Graviton Propagator}

The quantization of gravitons is in many respects similar to the massless scalar
case. Indeed the two physical polarizations of the graviton satisfy the
same wave equation as the massless scalar.
The issues with IR divergences are therefore similar.

To preserve the gauge invariance, it is preferable to work with the Lagrangian formalism
and impose a gauge fixing condition. A suitable gauge fixing
condition was found in~\cite{deVega:1998ia}. Following them,
we define
\bea
\gamma_{\mu \nu} dx^\mu dx^\nu &=& {1\over H^2 \t^2} (-d\t^2 + d\overrightarrow{x}^2)
\nonumber\\
g_{\mu \nu } &\equiv&  \gamma_{\mu \nu} +h_{\mu \nu}
\nonumber\\
\psi^\nu _\mu &=& h^\nu _\mu-{1\over 2}\d ^\nu _\mu h_{\lambda}^{\lambda}
\eea
where $\gamma_{\mu \nu}$ is the unperturbed metric of de Sitter space and
$h_{\mu \nu}$ is the metric perturbation.  Indices are raised and lowered
with the metric $\gamma_{\mu \nu}$, at lowest order.
We then impose the gauge-fixing condition
\bea
D_\nu \psi^\nu _\mu=2H^2\tau \psi^0 _\mu\label{gaugefix}
\eea
and find that the equations of motion for
the quadratic part of the action reduce to
\bea
\left(\square+{2\over \tau }\del_\tau \right) \chi_{ij}&=&0
\nonumber\\
\square\left({1\over \tau }\chi_{0i}\right)&=&0
\nonumber\\
\square\left({1\over \tau }\chi\right)&=&0
\eea
where $\chi_{00}=-\psi_0^0$, $\chi_{0i}=-\psi_i^0$, $\chi_{ij}=\psi_i^j$ and
$\chi=\chi_{11}+\chi_{22}+\chi_{33}+\chi_{00}$.

The $\chi_{ij}$ therefore obey the same wave equation as massless scalars, and
their propagators will be of the form
(\ref{modifiedprop}) with a different constant shift to the
zero-mode for each of the $\chi_{ij}$.
These shifts are related to the gauge symmetry; the gauge-fixing condition
(\ref{gaugefix}) imposes the constraints
\bea
-\partial_\tau \chi_{00} +\partial_i \chi_{0i}
+\tau^{-1} \chi &=& 0
\nonumber\\
-\partial_\tau \chi_{0i} +\partial_j \chi_{ji}
+2\tau^{-1} \chi_{0i} &=& 0
\eea
which can be used to fix $\chi_{00}$ and $\chi_{0i}$.
However, there is a residual gauge symmetry which is not fixed by
the above constraints.  The residual gauge transformations can be used
to generate a constant shift of the graviton perturbations (in fact, the IR divergence
 occurs because the gauge fixing term (\ref{gaugefix}) does not
completely fix the gauge).  The correlator can be made finite
 by an appropriate gauge choice which fixes the shifts, yielding
 \bea
G_{0,ijmn}^{\rm fin}(k,\t,\t') &=& \langle\chi_{ij}(k) \chi_{mn}(k)\rangle
\nonumber\\
&=& (g_{im}g_{jn}+g_{in} g_{jm}) \left(\lim_{\mu \rightarrow 0}
{H^2\over k^3}(1+ik\tau)(1-ik\tau')
e^{ik(\tau-\tau')}\theta(k-\mu)-\left(4\pi H^2 \ln {k_{IR} \over \mu} \right)\d^3(\vec{k})\right)
\label{gravitonprop}
\eea
which has no IR divergence.  The propagators for $\chi_{ij}$, $\chi_{0i}$, and $\chi_{00}$ are thus
free of IR divergences\footnote{It has been suggested~\cite{Gazeau:1999mi, GomezVergel:2007fd} that Gupta-Bleuler quantization
of the graviton also yields
unambiguous finite results for observables in de
Sitter space.}. Furthermore, the  ambiguity arising from the choice of $k_{IR}$ will not be present in
any gauge invariant expression, since the value of $k_{IR}$ can be shifted by a gauge transformation.

\section{Conclusion}

We have constructed infrared finite propagators
in inflating spacetimes by taking the usual Bunch-Davies propagators
and modifying the zero mode; a procedure we dubbed zero mode modification.
We do
not implicitly assume that all scalar fields are massive, or that inflation
is not eternal.
Instead, this method utilizes
the fact that in de Sitter spacetime, there is an inherent ambiguity in
boundary conditions for the zero mode of the Feynman propagator.  This
ambiguity can  be chosen to cancel the divergences which usually
arise from the integration of low-lying modes in a loop diagram.
This leaves over a finite ambiguity in the propagator, which can only
be set by observation.

Physically, this modification corresponds to the choice of a state
in which the variance of scalar field is finite at finite conformal time.
In practice, the only
observable effect of this choice is to change the variance (and in general,
higher zero-mode moments) of a stochastic field, and the result is to
divorce the moments of the field from the history of inflation before the era when
observable modes froze out.
The result is a coherent justification of the practical approach which
has often been taking when dealing with loop diagrams in the quantum
expansion of inflationary theories.
At leading order, one can see that calculations performed using
zero mode modification yield results similar to those obtained from
a hard IR cutoff at some momentum $k_{IR}$,  but these approaches will
differ by terms which are subleading  in $k_{IR}$.
Note that at a fundamental level, our message is that the description
of a massless scalar field in de Sitter space or its Poincar\'e patch depends on a
scale $H$ \emph{and} an extra arbitrary parameter (in the form of $k_{IR}$ here).

We have moreover argued that this approach can also be used to
calculate graviton loop diagrams, in which case the ambiguity in the
propagator is in fact a gauge ambiguity.  In a more detailed calculation
in a theory where gravity is dynamical, one should be able to see that, unlike
the case of the scalar propagator,
observables are independent of this ambiguity in graviton propagator.  It would
be interesting to address this issue with a concrete calculation of a gravitational
observable.

{\bf Acknowledgments} We are grateful to Niayesh Afshordi, Cliff Burgess, Richard Holman, David Seery,
Sarah Shandera and Andrew Tolley for useful conversations.
J.K. is grateful to the Perimeter Institute for its hospitality.
This research has been supported in part by funds from
the Natural Sciences and Engineering Research Council
(NSERC) of Canada, and Perimeter Institute. Research at the
Perimeter Institute is supported in part by the Government of
Canada through NSERC and by the Province of Ontario through the
Ministry of Research and Information (MRI). A.R. is supported in part by NSF Grant No. PHY-0653656.

\end{document}